\documentclass[9pt, onecolumn]{article}

\usepackage[margin=1in]{geometry}  

\usepackage{url}
\usepackage{hyperref}
\usepackage{pifont}
\usepackage{acronym}
\usepackage{multirow}
\usepackage{subcaption}
\usepackage{fontawesome5}
\usepackage{mdframed}
\usepackage[ruled,vlined,linesnumbered]{algorithm2e}
\usepackage{amsmath,amsfonts}
\usepackage{array}
\usepackage[table]{xcolor}
\usepackage{siunitx}
\usepackage{booktabs, makecell}
\newcommand{\cmark}{\ding{51}}
\usepackage{enumerate}
\usepackage{enumitem}
\usepackage{fancyhdr}
\usepackage{threeparttable}
\usepackage{newtxtext,newtxmath}
\newcommand{\VLLAMMBench}{\textsc{ViLLA-MMBench}}
\usepackage{pifont} 
\usepackage{tcolorbox}
\AtBeginDocument{%
  }

\acrodef{VOD}{Video on Demand}
\acrodef{FPS}{Frames per Second}
\acrodef{RS}{Recommender System}
\acrodef{LLM}{Large Language Model}
\acrodef{CF}{Collaborative Filtering}
\acrodef{CNN}{Convolutional Neural Network}
\acrodef{PCA}{Principal Component Analysis}
\acrodef{RAG}{Retrieval-Augmented Generation}
\acrodef{CCA}{Canonical Correlation Analysis}

\newcommand{\ali}[1]{\textcolor{olive}{{/* #1 (Ali) */}}}

\newcommand{\athena}[1]{\textcolor{blue}{{/* #1 (Athena) */}}}

\definecolor{red}{HTML}{fd8f8f}
\definecolor{greend}{HTML}{57e377}
\definecolor{greenl}{HTML}{b8fb8a}
\definecolor{yellow}{HTML}{fefdb4}

\colorlet{red}{red!50}
\colorlet{yellow}{yellow!50}
\colorlet{greenl}{greenl!50}
\colorlet{greend}{greend!50}

\begin{document}

\title{\VLLAMMBench: A Unified Benchmark Suite for LLM-Augmented Multimodal Movie Recommendation}

\author{
    Fatemeh Nazary$^{1}$, 
    Ali Tourani$^{2}$, 
    Yashar Deldjoo$^{1,*}$, 
    Tommaso Di Noia$^{1}$ \\[1ex]
    \small $^{1}$Department of Electrical and Information Engineering, Polytechnic University of Bari, Bari, Italy \\
    \small $^{2}$Interdisciplinary Centre for Security, Reliability, and Trust (SnT), University of Luxembourg, Luxembourg \\[1ex]
    \small \texttt{\{fatemeh.nazary, tommaso.dinoia\}@poliba.it, deldjooy@acm.org, ali.tourani@uni.lu} \\[1ex]
    \small $^{*}$Corresponding author
}

\maketitle

\begin{abstract}
Recommending long‑form video content requires an integrated treatment of visual, audio and textual modalities, yet most benchmarks focus on either raw item features or narrow fusion pipelines. We introduce \textbf{\VLLAMMBench}, a fully reproducible, extensible benchmark suite for next-generation LLM-augmented multimodal movie recommendation research. The toolkit leverages the widely-used MovieLens and MMTF-14K datasets, integrating and aligning item-level dense embeddings from three modalities: audio (block-level features and i-vector), visual (CNN and AVF), and text. Notably, it automatically augments missing or sparse item metadata using state-of-the-art Large Language Models (LLMs), such as OpenAI GPT (via the Ada model), generating high-quality synopses for thousands of movies. All text, whether raw or LLM-augmented, is embedded using configurable dense encoders, producing multiple ready-to-use sets (OpenAI Ada, LLaMA-2, Sentence-T5).

Furthermore, the pipeline in  \textbf{\VLLAMMBench} supports interchangeable \textbf{early}-, \textbf{mid}-, and \textbf{late}-fusion operators (concatenation, PCA, CCA, and rank-aggregation), and exposes a variety of backbone recommenders (MF, VAECF, VBPR, AMR, VMF) for ablation studies. All experimental parameters—including dataset splits, modality variants, fusion strategy, and LLM type—are declaratively specified via a single YAML file for transparent, versioned experimentation. Evaluation is comprehensive, covering not only accuracy (Recall, nDCG), but also beyond-accuracy axes: cold-start rate, coverage, novelty, diversity, and fairness, supporting rigorous, multi-metric benchmarking.

Experiments demonstrate that LLM-based text augmentation and dense embedding extraction directly benefit cold-start and coverage performance, especially when strong textual representations are fused with audio-visual descriptors. Systematic benchmarking reveals which embedding and fusion combinations are universal (strong across models) versus backbone- or metric-specific. Overall, the open-source code, embeddings, and configuration templates make it a robust foundation for reproducible, extensible, and fair comparison in multimodal recommender systems, and offer a clear step forward toward principled integration of generative AI in large-scale movie recommendation. All resources are publicly available at \url{https://recsys-lab.github.io/ViLLA-MMBench}.
\end{abstract}


\section{Introduction}\label{sec:introduction}

Recommending long-form video content remains a challenging task, despite recent advances in computer vision, audio processing, and large language models (LLMs). Movies and series deliver rich visual, auditory, and textual cues that need to be appropriately aligned and thereby integrated into a coherent representation before relevant recommendations can be produced.  Traditional collaborative filtering completely ignores item content, while many multimodal recommender systems rely on a single fusion strategy (typically simple feature concatenation) and offer limited transparency or reproducibility. In addition, video-oriented datasets are difficult to share because of copyright restrictions, and widely used benchmarks such as MovieLens and MMTF-14K provide only raw features or partial multimodal alignment~\cite{mmtf14k,movielens,fan2018movie}.

\vspace{0.25em}
The motivation for this work arises from several persistent obstacles:
\begin{itemize}
  \item[\faImage] \textbf{Scarcity of shareable content.} Full-length, high-quality movies span hours and are subject to strict copyright restrictions, severely limiting the availability of large-scale, publicly shareable datasets for reproducible research~\cite{mmtf14k, microlens, youtube8m}.
  \item[\faClock] \textbf{Temporal complexity.} Videos are inherently temporal, consisting of thousands of frames and variable-length audio, requiring summarization into compact representations that preserve semantics, narrative, and affect~\cite{mmrec, mmsl, microlens, zhou2020s3}. Yet, temporal aggregation remains underexplored in multimodal fusion pipelines.
  \item[\faProjectDiagram] \textbf{Fusion strategy uncertainty.} There is no consensus on optimal \emph{fusion} strategies, as highlighted in recent surveys and benchmarks~\cite{ducho_elliot, mmrec, microlens}. Pipelines often default to early-fusion (feature concatenation, as in MMRec-LLM \cite{mmrec_llm}, or Ducho~\cite{ducho2}), mid-fusion with learned projections (PCA, CCA~\cite{mmrec}), often with less attention to system-level fusion. The reproducibility and interpretability of these choices remain open issues.
  \item[\faLightbulb] \textbf{The rise of LLM-driven augmentation.} Recent LLMs—such as GPT-4, LLaMA, and LVLMs—have dramatically improved the generation and embedding of textual side information~\cite{recgpt, mmrec_llm, llmrec}. LLMs can synthesize fluent synopses for items with sparse or missing metadata (addressing the long-tail problem in MovieLens~\cite{movielens, mmrec_llm}), and their embeddings encode broad world knowledge~\cite{recgpt}. However, principled strategies for fusing LLM-generated signals with audio-visual descriptors and for evaluating their impact on user-facing and beyond-accuracy metrics are still lacking~\cite{mmrec_llm, mmrec}.
\end{itemize}

Recent research has nevertheless opened two significant opportunities. First, compact visual and audio embeddings can be extracted from trailers via pre-trained convolutional and audio models, enabling efficient content representation. Second, LLMs can fill metadata gaps, thereby enriching textual features. Yet, existing benchmarks either focus on general multimodal frameworks without LLM integration (e.g., Ducho, MMRec) or propose LLM-augmented synopses without a unified evaluation pipeline (\S \ref{sec_sota}). Consequently, the community still lacks an open benchmark that simultaneously:

\begin{enumerate}[]
  \item integrates dense audio, visual, and LLM-generated textual descriptors,
  \item exposes interchangeable early-, mid-, and late-fusion operators,
  \item supports diverse recommendation backbones, and
  \item reports a comprehensive suite of accuracy and beyond-accuracy metrics.
\end{enumerate}

\medskip
\textbf{Contributions.} This paper introduces \emph{ViLLA-MMBench}, a unified benchmark suite for \emph{LLM-augmented multimodal movie recommendation}. In contrast to earlier prototypes, we provide a complete Python package that can be installed via \texttt{pip}, configured through a YAML file and executed on local machines or cloud servers. Our contributions are fourfold:
\begin{itemize}
    \item \textbf{Unified multimodal pipeline.} ViLLA-MMBench aligns audio, visual and textual embeddings for MovieLens-1M and MMTF-14K, augments missing synopses with a variety of LLMs, and supplies ready-to-use dense text embeddings produced by OpenAI-Ada, Sentence-T5 and LLaMA-2. This results in a coherent tri-modal representation for roughly \num{1000} movies after modality filtering.
    \item \textbf{Configurable fusion strategies.} The toolkit supports early-fusion methods (concatenation, PCA, CCA), mid-fusion (projected representations), and late-fusion (ensemble ranking). Because each modality is loaded through a dedicated module, new embeddings or fusion techniques can be incorporated with minimal engineering effort.
    \item \textbf{Diverse recommendation backbones and beyond-accuracy metrics.} We implement matrix factorization, variational autoencoder collaborative filtering, and content-aware models such as VBPR, VMF and AMR, and we expose a simple interface to add more algorithms. A grid-search module performs GPU-aware hyper-parameter optimisation and evaluates models on an extensive suite of metrics, including Recall, nDCG, coverage, cold-start rate, novelty, intra-list diversity, and calibration bias.
    \item \textbf{Reproducibility and extensibility.} All experimental parameters (dataset, split strategy, fusion operator, LLM choice, modality variant) are specified in a YAML configuration file. The codebase contains modular loaders for data and embeddings, model training and evaluation, and utility functions, thereby facilitating the integration of new datasets or modalities without altering the core pipeline.
\end{itemize}
By providing code, documentation and pre-processed embeddings, ViLLA-MMBench serves as a plug-and-play platform for systematic research into LLM-augmented multimodal recommendation, addressing limitations in existing work and enabling controlled ablation studies across fusion operators, modalities and recommendation models.

\begin{table*}[!t]
    \centering
    \fontsize{6.2}{7.5}\selectfont
    \setlength{\tabcolsep}{3.5pt}
    \begin{threeparttable}
    \caption{Multimodal movie/video recommendation systems and resources. \textbf{Video Type}: Tr = Trailer, µV = Micro-video. \textbf{Modalities}: icons for Visual (\faImage), Audio (\faMusic), Text (\faFont). \textbf{Fusion}: both timing stage and technique. \textbf{LLM}: \cmark = LLM augmentation. \textbf{RS Model}: core recommender family; Model Fusion indicates late fusion of models or rank aggregation.}
    \begin{tabular}{c|c|p{1.8cm}|c|c|c|c|c|p{3.8cm}|c}
    \toprule
        \multirow{2}{*}{\textbf{Class}} &
        \multirow{2}{*}{\textbf{Type}} & \multirow{2}{*}{\textbf{System}} &
        \multicolumn{2}{c|}{\textbf{Modalities}} & \multicolumn{2}{c|}{\textbf{Fusion}} &
        \textbf{RS Model} & \multirow{2}{*}{\textbf{Key Insight}} & \multirow{2}{*}{\textbf{Link}} \\
    \cmidrule(lr){4-5}\cmidrule(lr){6-7}
        & & &
        \tiny\makecell{\faImage\\\faMusic\\\faFont} &
        \makecell{LLM\\\cmark, type} &
        Stage & 
        Type & 
        \makecell{Family\\Model Fusion} & & \\
    \midrule
    
        \multirow{16}{*}{\rotatebox{90}{\shortstack{General-Purpose Multi-Modal Recommenders}}}
        & -- & Ducho 2.0 \cite{ducho2} &
        \tiny\makecell{\faImage\\\faMusic\\\faFont} &
        -- &
        Early &
        \makecell{Basic \\(Concat, sum, mean etc.)} 
         &
        \makecell{VBPR, BM3, FREEDOM\\--} & 
        Turns raw V/A/T into embeddings; user plugs any RS model. 
        & \href{https://github.com/sisinflab/Ducho}{[code]} \\ \cline{2-10}
    
        & -- & Ducho$\times$Elliot \cite{ducho_elliot} &
        \tiny\makecell{\faImage\\\faMusic\\\faFont} &
        -- &
        Early &
        \makecell{Basic \\(Concat, sum, mean etc.)} &
        \makecell{12 models \\--} &
        Combines features in Ducho with evaluation in Elliot for reproducible benchmarks. 
        & \href{https://github.com/sisinflab/Ducho-meets-Elliot}{[code]} \\ \cline{2-10}
    
        & -- & Rec-GPT4V \cite{recgpt} &
        \tiny\makecell{\faImage\\\faFont} &
        \cmark\,LVLM&
        \makecell{Late}&
         \makecell{LLM-based \\ (prompt level)} &
        \makecell{LLM\\--} &
        LVLM “see-and-chat” recommendations\footnote{“see-and-chat” refers to prompting an LVLM with both image (video frames, posters) and text (user history, metadata, title, etc.), and receiving a text-based answer (“chat”)—here, a recommendation list or justification.}—no additional training needed. & - \\ \cline{2-10}
    
        & – & MMSSL \cite{mmsl} &
        \tiny\makecell{\faImage\\\faMusic\\\faFont} &
        -- &
        Hybrid &
               \makecell{Attention}
        &
        \makecell{Deep (GNN-based CF)\\--} &
        Claims SSL “alleviates label sparsity” and “integrates unlabeled data” at the cost of heavier training. Training complexity is not directly measured. 
         & \href{https://github.com/HKUDS/MMSSL}{[code]} \\ \cline{2-10}
    
        & – & MMRec \cite{mmrec} &
        \tiny\makecell{\faImage\\\faMusic\\\faFont} &
        -- &
        Early &
        Concat, PCA, Attn &
        \makecell{CF, Deep (NGCF, VBPR, etc. ); \\--} &
        Toolbox unifying many multimodal RS; ideal for surveys and ablations.
        & \href{https://github.com/enoche/MMRec}{[code]} \\ \cline{2-10}
    
        & – & MMRec-LLM \cite{mmrec_llm} & 
        \tiny\makecell{\faImage\\\faFont} &
        \cmark\,LLM &
        Early &
        \makecell{Basic \\ (Concat)} &
        \makecell{CF, Deep\\--} &
       Shows that LLM-generated synopses (fusing visual + text cues) yield improved recommendation performance. & - \\

    \midrule
    \midrule
    
        \multirow{9}{*}{\rotatebox{90}{\shortstack{Video Recommender}}} & µV & MicroLens \cite{microlens} &
        \tiny\makecell{\faImage\\\faMusic\\\faFont} &
        -- &
        Early &
        \makecell{Basic \\(Concat)} &
        \makecell{–- \\ --} &
        Billion-interaction micro-video corpus—great for deep seq-RS research.
        & \href{https://github.com/westlake-repl/MicroLens}{[data]} \\ \cline{2-10}
    
    
        & Tr & MMTF-14K \cite{mmtf14k} &
        \tiny\makecell{\faImage\\\faMusic\\\faFont} &
        -- &
        - &
        – &
        \makecell{–- \\Late (Rank Aggregation)} &
        Staple trailer-based multimodal dataset aligned with MovieLens.
        & \href{https://zenodo.org/records/1225406}{[data]} \\ \cline{2-10}
        & Tr & \makecell{ \\ \textbf{Ours}} &
    \tiny\makecell{\\ \\ \faImage\\\faMusic\\\faFont} &
        \cmark\,LLM &
        Hybrid &
        \makecell{\shortstack{Early \\ (Concat, PCA, CCA)}} &
        \makecell{VBPR, AMR, VMF\\Late (Rank Aggregation)} &
        Fully reproducible, plug-and-play platform for LLM-enriched, multimodal recommendations aligned with MovieLens. \textbf{Multiple fusion techniques are employed both at the feature level and system level.}
        & \href{https://recsys-lab.github.io/ViLLA-MMBench}{[code]} \\
    \bottomrule
    
    \end{tabular}
    \label{tab:mm_rs_compact}
\begin{tablenotes}[flushleft]
\footnotesize
\item[1] Rec-GPT4V: \textit{“See-and-chat” refers to prompting an LVLM with both image (video frames, posters) and text (user history, metadata, title, etc.), and receiving a text-based answer (“chat”)—here, a recommendation list or justification.}
\item[2] Rec-GPT4V: \textit{\textbf{Late fusion} (“prompt-level” after all modalities are presented).}
\item[3] MMSSL: \textit{\textbf{Hybrid fusion:} Features are fused at both early (representation) and late (joint learning) stages. It uses cross-modal self-attention and MM graph attention.}
\item[4] MMRec-LLM: \textit{Uses GPT-3.5 to generate “synopsis” (synthetic) text for items, combining image tags + text.}
\end{tablenotes}
\end{threeparttable}
\end{table*}

\begin{itemize}
    \item \textbf{Unified multimodal pipeline:} \VLLAMMBench~natively ingests and aligns audio, visual, and textual embeddings, integrating metadata and dense content features from MovieLens 1M~\cite{movielens}, MMTF-14K~\cite{mmtf14k}, and in-house LLM-augmented synopses~\cite{mmrec_llm} produced in this work.
    \item \textbf{Configurable fusion strategies:} The toolkit provides interchangeable early (concatenation, PCA, CCA), mid, and late/system-level (ensemble ranking) fusion methods, enabling controlled ablation studies and benchmarking to advance research in this area with respect to recent advances~\cite{mmrec, mmrec_llm, ducho2, ducho_elliot}.
    \item \textbf{LLM-based augmentation:} We auto-generate rich, human-readable synopses for MovieLens movies lacking metadata, and provide multiple ready-to-use embedding sets (Sentence-T5, LLaMA-2, OpenAI Ada, etc.)—each on both raw and LLM-augmented synopses.
    \item \textbf{Systematic evaluation:} The framework benchmarks state-of-the-art recommenders—including MF, VAECF, VBPR, AMR, hybrids, and recent GNNs—under GPU-aware hyperparameter grids, reporting more than ten metrics covering accuracy (Recall, nDCG), coverage, cold-start, fairness, novelty, and diversity.
    \item \textbf{Reproducibility and extensibility:} Every configuration, split, and metric is declarative and versioned; new modalities or models can be incorporated via drop-in loaders or subclassing, following best practices for transparent, reproducible research~\cite{ducho_elliot, microlens, mmrec_llm}.
\end{itemize}

By making all code and resources publicly available\footnote{\url{https://recsys-lab.github.io/ViLLA-MMBench}}, \VLLAMMBench~provides a robust, plug-and-play platform for systematic research on multimodal, LLM-enriched video recommendation. Our toolkit lays the groundwork for reproducible, extensible benchmarking and fair comparison across fusion operators, model classes, and evaluation criteria—addressing the key open questions identified in prior surveys and recent benchmarks~\cite{mmrec, mmrec_llm, recgpt, ducho_elliot}, and highlighted in Table~\ref{tab:mm_rs_compact} (\S\ref{sec_sota}).

\vspace{0.5em}
In summary, this work contributes the first unified, fully reproducible resource for exploring how LLM-augmented text, audio, and vision interact in large-scale movie recommendation—a crucial step toward the next generation of multimodal recommender systems.

\subsection{Related Multimodal Frameworks and Gaps}
\label{sec_sota}

Over the past decade, a variety of multimodal recommender systems have emerged, each supporting different modalities and fusion strategies at varying levels of scale and reproducibility. Table~\ref{tab:mm_rs_compact} organizes these systems and resources by their \textit{modality support}, fusion strategies, use of large language models (LLMs), system-level fusion, and recommendation model families.

\paragraph{General-purpose frameworks.}  
Systems such as  Ducho 2.0 \cite{ducho2}, and MMRec \cite{mmrec} primarily leverage early fusion through concatenation of audio, visual, and textual features, positioning themselves as flexible frameworks or toolkits for rapid benchmarking and ablation studies. More specialized systems like Rec-GPT4V \cite{recgpt} use LLM-based visual-to-text capabilities, enabling “see-and-chat” style recommendations without retraining, whereas MMRec-LLM \cite{mmrec_llm} integrates synopsis generation via LLMs to significantly enhance side-information quality and recommendation accuracy. All of them accept video, audio, and text, yet they differ markedly in how they marry these signals: early concatenation is still the dominant strategy (Cornac, Ducho), but attention mechanisms (MMSSL \cite{mmsl}) and configurable PCA/Attn hybrids (MMRec) appear when scalability or interpretability becomes an issue. 

\paragraph{Video-centered resources.}  
Datasets such as MicroLens~\cite{microlens}, MMTF-14K~\cite{mmtf14k} have become indispensable for developing scalable and realistic benchmarks, supporting research on sequence modelling, temporal aggregation, and modality alignment in video recommendation. However, they typically offer only raw interaction logs or extracted features and lack built-in pipelines for systematic fusion, evaluation, or LLM-augmented descriptors.

\paragraph{Gaps and limitations.}  
Despite these advances, several gaps remain: most systems still default to basic early fusion (e.g., concatenation), and only a handful support hybrid or late/system-level fusion through modular pipelines~\cite{mmrec, mmsl, mmrec_llm}. The integration of LLM-based features is not yet standardized; benchmarks often lack fair, flexible support for modality selection, multi-fusion strategies, and beyond-accuracy evaluation criteria (e.g., fairness, diversity, cold-start, coverage)~\cite{microlens, mmrec, ducho_elliot, mmrec_llm, zhou2020s3}. Temporal aspects—so critical in video—are often only superficially addressed or handled outside the main fusion framework.

\paragraph{\VLLAMMBench.}  
To address these limitations, \textbf{\VLLAMMBench} provides the first open-source, fully reproducible pipeline for audio-visual-textual video recommendation with native LLM support. Compared to prior work, our proposed systems puts forward the following novel steps: (i) unifying feature extraction and LLM-driven augmentation for all MovieLens/MMTF-14K items; (ii) offering fully configurable early, mid, and late fusion operators; (iii) exposing a plug-and-play layer for integrating new models or evaluation metrics; and (iv) supporting comprehensive benchmarking across accuracy and beyond-accuracy axes. Our pipeline is designed for extensibility, declarative experimentation, and fair, apples-to-apples comparison—closing the reproducibility, flexibility, and LLM-integration gaps identified in Table~\ref{tab:mm_rs_compact}.

\section{Technical Background}
\label{sec:background}

We organize the technical background and related models into three principal categories: (i) \emph{interaction-only collaborative filtering}, (ii) collaborative filtering models that incorporate side information, and (iii) system-level fusion strategies that aggregate outputs from independently trained recommenders. Throughout this section, we consistently define all variables and formal notation to ensure clarity and coherence.

\subsection{Compared Recommendation Models}

\noindent \textbf{Notation and Problem Setting.} Let $\mathcal{U} = \{1, \dots, |\mathcal{U}|\}$ denote the set of users, and $\mathcal{I} = \{1, \dots, |\mathcal{I}|\}$ the set of items (e.g., movies). In the implicit feedback scenario, interactions are captured by the set
\begin{equation}
  \mathcal{R} = \left\{ (u, i) ~\middle|~ u \in \mathcal{U},~ i \in \mathcal{I},~ r_{ui} = 1 \right\},
  \label{eq:R}
\end{equation}
where $r_{ui} \in \{0, 1\}$ indicates whether user $u$ has interacted positively with item $i$. Unobserved pairs $(u, i) \notin \mathcal{R}$ may correspond to either uninterest or lack of exposure.

The goal of a recommender system is to learn a scoring function $\hat{r}: \mathcal{U} \times \mathcal{I} \rightarrow \mathbb{R}$ that estimates the affinity of user $u$ for the item $i$. For each user $u$, items are ranked in descending order of $\hat{r}_{ui}$, producing personalized top-$N$ recommendations.\\

\noindent
\textbf{Interaction-Only Baselines (Pure CF).} As baselines for our multimodal recommender models and as building blocks for ensemble-based fusion, we employ the following two models 

\paragraph{Matrix Factorization (MF)~\cite{mf}.}
Matrix Factorization represents each user $u$ and item $i$ by latent vectors $\mathbf{p}_u, \mathbf{q}_i \in \mathbb{R}^d$ in a shared $d$-dimensional space, with global and individual bias terms:
\begin{equation}
  \hat{r}_{ui} = \mu + b_u + b_i + \mathbf{p}_u^\top \mathbf{q}_i,
  \label{eq:mf}
\end{equation}
where $\mu$ is the global bias, $b_u$, $b_i$ are user and item biases, respectively. The model parameters are learned by minimizing the regularized squared error:
\begin{equation}
  \min_{\mathbf{P}, \mathbf{Q}, b, \mu} \sum_{(u, i) \in \mathcal{R}} (r_{ui} - \hat{r}_{ui})^2 + \lambda \left( \|\mathbf{p}_u\|_2^2 + \|\mathbf{q}_i\|_2^2 \right),
  \label{eq:mf-loss}
\end{equation}
where $\lambda > 0$ is the regularization coefficient.

\begin{table*}[!t]
  \centering
  \caption{Deterministic fusion operators used in this work.}
  \label{tab:fusion-ops}
  \begin{tabular}{@{}lll@{}}
    \toprule
    \textbf{Tag} & \textbf{Operator \(f(\mathbf e_i^{(\mathrm{aud})}, \mathbf e_i^{(\mathrm{vis})}, \mathbf e_i^{(\mathrm{txt})})\)} & \textbf{Output Dim. \(d_f\)} \\
    \midrule
    \texttt{concat} & \( \mathbf e_i = [\mathbf e_i^{(\mathrm{aud})};\, \mathbf e_i^{(\mathrm{vis})};\, \mathbf e_i^{(\mathrm{txt})}] \) & \( d_{\mathrm{aud}} + d_{\mathrm{vis}} + d_{\mathrm{txt}} \) \\

    \texttt{pca\_\(\rho\)} & \( \tilde{\mathbf e}_i = \mathbf P^\top\,\mathrm{zscore}(\mathbf e_i^{\text{concat}}) \), retain \( d_\rho \) s.t.\ \( \sum_{j=1}^{d_\rho}\lambda_j / \sum_j \lambda_j \geq \rho \) & \( d_\rho \) \\

    \texttt{cca\_\(k\)} & \( \tilde{\mathbf e}_i = [\mathbf W_1^\top \mathbf e_i^{(1)}]_{1:k} \), \( \mathbf W_1, \mathbf W_2 \) maximize \( \mathrm{corr}(\mathbf W_1^\top \mathbf e_i^{(1)}, \mathbf W_2^\top \mathbf e_i^{(2)}) \) & \( k \) \\
    \bottomrule
  \end{tabular}
\end{table*}

\paragraph{Variational Autoencoder for Collaborative Filtering (VAECF)~\cite{vaecf}.}
VAECF encodes each user’s interaction vector $\mathbf{x}_u \in \{0, 1\}^{|\mathcal{I}|}$ into a Gaussian latent code $q_\phi(\mathbf{z}_u|\mathbf{x}_u)$ and reconstructs it via a decoder $p_\theta(\mathbf{x}_u|\mathbf{z}_u)$. Learning maximizes the evidence lower bound (ELBO):
\begin{equation}
  \mathrm{ELBO}(\theta, \phi) = \sum_{u \in \mathcal{U}} \Big[
    \mathbb{E}_{q_\phi} \left[ \ln p_\theta(\mathbf{x}_u|\mathbf{z}_u) \right] - \beta\, \mathrm{KL}\left(q_\phi \,\|\, p\right)
  \Big],
  \label{eq:vaecf}
\end{equation}
where $\beta$ controls the regularization strength and $\mathrm{KL}(\cdot\|\cdot)$ denotes the Kullback–Leibler divergence.

\noindent
\textbf{Collaborative Filtering with Side Information.} Many recent models incorporate item side information (e.g., text, image, audio) to address cold-start and improve generalization. We distinguish two principal approaches to model textual and multimodal signals:
\begin{itemize}
    \item[(a)] \textbf{Raw Text Models}: These methods (e.g., HFT~\cite{hft}, CDL~\cite{cdl}) operate directly on raw text using topic models or neural networks to extract interpretable item representations.
    \item[(b)] \textbf{Dense Embedding Models}: These approaches leverage precomputed dense vectors derived from deep neural encoders, applicable to text, audio, and visual modalities. They enable efficient and flexible integration of rich semantic cues into collaborative filtering.
\end{itemize}

Given our goal of systematically evaluating the effect of aligned dense embeddings for audio, visual, and textual modalities, we focus on models—VBPR, VMF,\footnote{Due to the extensive experiments and the superior performance of VBPR and AMR, for space limitations, we focus our report on these two models; complete results—including VMF and concatenation-based fusion which were omitted here (see \S \ref{subsec:mm-fusion})—are available on our GitHub.} and AMR—designed for direct embedding input, and previously validated for visual, multimedia, and textual recommendation tasks. This ensures fair and balanced comparison across modalities and models.

Formally, for each item $i \in \mathcal{I}$, let:
\begin{itemize}
    \item $\mathbf{e}_i^{(\mathrm{txt})} \in \mathbb{R}^{d_\mathrm{txt}}$: $\ell_2$-normalized text embedding
    \item $\mathbf{e}_i^{(\mathrm{vis})} \in \mathbb{R}^{d_\mathrm{vis}}$: $\ell_2$-normalized visual embedding
    \item $\mathbf{e}_i^{(\mathrm{aud})} \in \mathbb{R}^{d_\mathrm{aud}}$: $\ell_2$-normalized audio embedding
\end{itemize}
These can be concatenated as $\mathbf{e}_i = [\,\mathbf{e}_i^{(\mathrm{txt})} \,\|\, \mathbf{e}_i^{(\mathrm{vis})} \,\|\, \mathbf{e}_i^{(\mathrm{aud})}\,]$.

\paragraph{VBPR~\cite{vbpr}.}
VBPR extends Bayesian Personalized Ranking (BPR) to incorporate visual (or general content) features:
\begin{equation}
  \hat{r}_{ui} = \mathbf{p}_u^\top \mathbf{q}_i + \mathbf{w}_u^\top \mathbf{e}_i,
\end{equation}
where $\mathbf{w}_u$ captures user-specific preferences for side features.

\paragraph{VMF~\cite{vmf}.}
VMF is a multi-modal extension of MF, projecting the fused side information to the collaborative latent space:
\begin{equation}
  \mathbf{q}_i = \mathbf{H}\, \mathbf{e}_i, \quad \hat{r}_{ui} = \mu + b_u + b_i + \mathbf{p}_u^\top \mathbf{q}_i,
\end{equation}
with $\mathbf{H} \in \mathbb{R}^{d \times d_e}$ a learned projection matrix.

\paragraph{AMR~\cite{amr}.}
AMR uses a gating (attention) network $g(\cdot)$ to assign weights to each modality:
\begin{equation}
  \hat{r}_{ui} = \mathbf{p}_u^\top \mathbf{q}_i + g\!\bigl(\mathbf{e}_i^{(\mathrm{txt})},\, \mathbf{e}_i^{(\mathrm{vis})},\, \mathbf{e}_i^{(\mathrm{aud})}\bigr).
\end{equation}

\subsection{Multi-Modal Fusion of Embeddings}
\label{subsec:mm-fusion}

For every item \(i\in\mathcal I\) we pre-compute three
\emph{modality-specific} embeddings, all $\ell_2$-normalised:

\[
  \mathbf e_i^{(\mathrm{aud})}\in\mathbb R^{d_{\mathrm{aud}}},\qquad
  \mathbf e_i^{(\mathrm{vis})}\in\mathbb R^{d_{\mathrm{vis}}},\qquad
  \mathbf e_i^{(\mathrm{txt})}\in\mathbb R^{d_{\mathrm{txt}}}.
\]

A deterministic operator  
\(
  f:\mathbb R^{d_{\mathrm{aud}}}\!\times\!
    \mathbb R^{d_{\mathrm{vis}}}\!\times\!
    \mathbb R^{d_{\mathrm{txt}}}
    \to\mathbb R^{d_f}
\)
maps the triplet  
\(\bigl(\mathbf e_i^{(\mathrm{aud})},
        \mathbf e_i^{(\mathrm{vis})},
        \mathbf e_i^{(\mathrm{txt})}\bigr)\)
to a single multimodal descriptor
\(
  \mathbf e_i
  =f\!\bigl(\mathbf e_i^{(\mathrm{aud})},
            \mathbf e_i^{(\mathrm{vis})},
            \mathbf e_i^{(\mathrm{txt})}\bigr).
\)

We evaluate three early-fusion rules:

\begin{itemize}
  \item \textbf{Concatenation}:\;
        \(\displaystyle
          \mathbf e_i
          =\bigl[\mathbf e_i^{(\mathrm{aud})}\!;
                  \mathbf e_i^{(\mathrm{vis})}\!;
                  \mathbf e_i^{(\mathrm{txt})}\bigr]\),
        giving dimensionality
        \(d_f=d_{\mathrm{aud}}+d_{\mathrm{vis}}+d_{\mathrm{txt}}\).

  \item \textbf{Principal Component Analysis (PCA)}:\;
        form the concatenated vector above, standardise it
        (z-score), then project onto the first \(d_\rho\) principal
        components such that the retained cumulative variance
        satisfies
        \(\sum_{j=1}^{d_\rho}\lambda_j/\sum_j\lambda_j\ge\rho\).

  \item \textbf{Canonical Correlation Analysis (CCA)}:\;
        Split the concatenated vector in two equal halves,
        learn linear maps that maximise the correlation between
        the projected halves, and keep the first \(k\) canonical
        dimensions.
\end{itemize}

The resulting vector \(\mathbf e_i\) is passed to the downstream
recommender either as an \texttt{ImageModality} or a
\texttt{FeatureModality} (Cornac API), depending on the model. The three operators used in this work are summarised in
Table~\ref{tab:fusion-ops}. 

\subsection{System-Level Fusion of Recommender Outputs}
\label{subsec:sys-fusion}

Assume \(M\) independently trained recommenders indexed by
\(\mathcal S=\{1,\dots,M\}\).
For user \(u\) and system \(m\in\mathcal S\), let
\(
  L_{u}^{(m)}=(i_{u,1}^{(m)},i_{u,2}^{(m)},\dots)
\)
be its ranked list.
The aim is to merge
\(\{L_{u}^{(m)}\}_{m\in\mathcal S}\)
into a single meta-ranking
\(F_u=(i_{u,1},i_{u,2},\dots)\).
We apply four classical, parameter-free aggregation rules:

\begin{description}[leftmargin=1.5em, labelsep=0.5em]
  \item[Borda count.]
    Each system assigns
    \(s_{u,i}^{(m)}=|\mathcal I|-\mathrm{rank}_{L_u^{(m)}}(i)+1\);
    fused scores are
    \(s_{u,i}=\sum_{m}s_{u,i}^{(m)}\).

  \item[Weighted Borda.]
    Extends the above with weights
    \(w_m\) s.t.\ \(\sum_m w_m=1\):
    \(s_{u,i}=\sum_m w_m s_{u,i}^{(m)}\).

  \item[Average rank.]
    Computes
    \(\bar r_{u,i}=M^{-1}\sum_m\mathrm{rank}_{L_u^{(m)}}(i)\)
    and sorts by ascending \(\bar r_{u,i}\).

  \item[Reciprocal-rank fusion (RRF)~\cite{cormack2009reciprocal}.]
    Uses
    \(
      s_{u,i}=\sum_m \bigl(k+\mathrm{rank}_{L_u^{(m)}}(i)\bigr)^{-1}
    \)
    with \(k=60\) (fixed).
\end{description}


The meta-rankings $F_u$ are evaluated using the same top-$k$ accuracy metrics (Recall@10, nDCG@10, Hit Rate@10) and beyond-accuracy metrics (catalogue coverage, cold-start rate) as those used for individual recommenders.

\vspace{2pt}
\noindent

\begin{figure*}[!t]
  \centering
  \includegraphics[width=1.0\linewidth]{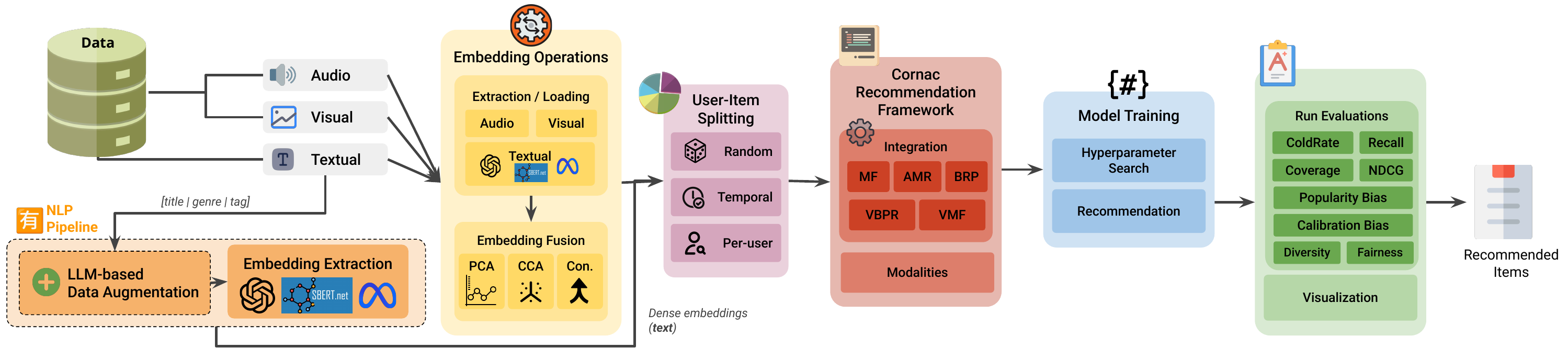}
  \caption{The architecture of the proposed toolkit for movie recommendation.}
  \label{fig_overal}
\end{figure*}

\section{ViLLA-MMBench Design and Data Pipeline}
In this section, we provide a detailed technical description of the ViLLA-MMBench implementation, covering the system overview (\S \ref{subsec:system_overview}), configuration and customization options (\S \ref{subsec:cofigs}), and the suite of evaluation metrics (\S \ref{subsec:metrics}). Each subsection includes comprehensive explanations and implementation details.

\subsection{System Overview}
\label{subsec:system_overview}

Figure~\ref{fig_overal} illustrates the architecture of ViLLA-MMBench, which is divided into four stages: data preparation, textual enrichment and embedding, multimodal alignment and fusion, and training and evaluation. After configuring the framework via the \texttt{config.yml} file, the entire data preparation and training pipeline can be executed sequentially by running the \texttt{main.py} script, which orchestrates all necessary procedures based on the specified settings.
\begin{enumerate}
     \item[\ding{182}] \textbf{Data preparation and ingestion.} The framework loads datasets through a uniform \texttt{pandas} interface. The \texttt{prepareML} function downloads and reads the MovieLens dataset (\texttt{100K} or \texttt{1M} variants, based on the given configuration), applies k-core filtering (if set), and performs train/test splitting according to the selected strategy (\texttt{random}, \texttt{temporal}, or \texttt{per\_user}). The \texttt{prepareModalities} function loads and preprocesses textual data from our in-house dataset, as well as visual and audio embeddings from the MMTF-14K dataset. The variants of these modalities to be loaded are also adjustable from the configuration file. While we provide loaders for MovieLens-1M and MMTF-14K, any dataset with user--item--rating triples can be ingested by implementing a similar loader. Any contribution for adding other modalities or datasets requires implementing simple loader functions by extending the modular structure in the \texttt{data} directory of the framework, similar to the existing \texttt{loadText}, \texttt{loadAudio}, or \texttt{loadVisual} functions within the respective \texttt{text.py}, \texttt{audio.py}, or \texttt{visual.py} files.
     \item[\ding{183}] \textbf{Textual enrichment and embedding.} For each item, a textual description is created by concatenating the title, genres, and tags or by prompting an LLM to produce a 100--150-word synopsis. The prompt and output are logged for transparency. The resulting text is embedded using the specified model (OpenAI-Ada, Sentence-T5, LLaMA-2), yielding a dense vector. Since the embedding code resides in \texttt{villa\_mmbench/data/text.py}, adding new LLMs or embedding models only requires registering a function in this module.
     \item[\ding{184}]\textbf{Multimodal alignment and fusion.} Audio, visual, and textual embeddings are aligned via item identifiers. Three deterministic early-fusion operators are supplied (\textsc{concat}, \textsc{pca}$_\rho$, or \textsc{cca}$_k$), representing concatenation, PCA (retaining a fixed proportion of variance) and CCA (projecting halves to maximise correlation). Mid- and late-fusion strategies are available via configuration. The \texttt{prepareModalities} function merges modalities, handles missing values and wraps the resulting features into the appropriate Cornac objects for downstream recommendation.
     \item[\ding{185}]\textbf{Training, evaluation and logging.} The \texttt{gridSearch} module in the framework's \texttt{grid.py} file performs hyper-parameter optimisation for the chosen model class, optionally using GPU resources. It evaluates each candidate on recall and nDCG and records the best configuration. Finally, the \texttt{generateLists} function in \texttt{processes.py} trains the selected model on the full training data and produces recommendation lists for each test user, computing metrics such as recall, nDCG, coverage, cold-start rate, novelty, intra-list diversity, popularity bias, and fairness. Results are saved as CSV files (by default in the \texttt{outputs} folder) for subsequent analysis. 

\end{enumerate}

\subsection{Configuration and Customization}
\label{subsec:cofigs}

Experiments are specified entirely through a YAML file (\texttt{config.yml}). A rich suite of parameters fully specifies an experiment.
\begin{itemize}[leftmargin=1.25em,itemsep=2pt]
  \item \textbf{General}: \texttt{root\_path}
  \item \textbf{Dataset \& Split}:
  \begin{itemize}
      \item \textbf{MovieLens}: \texttt{100k|1m}
      \item \textbf{Split}: \texttt{random|temporal|per\_user}
      \item  \textbf{Cold-start}: \texttt{k\_core}, \texttt{simulate\_cold\_start}
  \end{itemize}
  \item \textbf{Modality}:
  \begin{itemize}
      \item \textbf{LLM}: \texttt{openai|st|llama}
      \item \textbf{Augmentation}: \texttt{true|false}
      \item \textbf{Audio-variant}: \texttt{blf|i\_vec}
      \item \textbf{Visual-variant}: \texttt{cnn|avf}
  \item \textbf{Fusion}: \texttt{concat|pca\_$\rho$|cca\_k} (\S\ref{subsec:mm-fusion})
  \end{itemize}
  \item \textbf{Experiment}: \texttt{seed}, \texttt{epochs}, \texttt{use\_gpu}, \texttt{fast\_prototype}, \texttt{parallel\_hpo}, \textit{etc.}  
  \item \textbf{Recommendation and Experiments}: 
  \begin{itemize}
      \item \textbf{Model}: \texttt{cf|vbpr|amr|vmf}
      \item \textbf{Runtime}: \texttt{seed}, \texttt{epochs}, \texttt{gpu\_id}, \texttt{fast\_prototype}
  \end{itemize}
\end{itemize}

\noindent Default values reproduce all results reported in~\S\ref{sec_eval} and more information is provided below:

\begin{description}
    \item[Dataset and split:] Select a predefined \texttt{movielens} version, choose a splitting mode (\texttt{random}, \texttt{temporal}, \texttt{per\_user}), and set the test ratio and \texttt{k\_core}. Researchers can thus replicate experiments across different scenarios or apply the framework to new datasets.
    \item[Modality variants:] Choose audio embeddings (\texttt{blf} or \texttt{i\_ivec}), visual features (\texttt{cnn} or \texttt{avf}) and decide whether to use LLM-generated text. Adding additional modalities, such as user demographics or interaction contexts, is straightforward through the data modules.
\item[Fusion operator:] Specify \texttt{concat}, \texttt{pca\_}$\rho$, or \texttt{cca\_}$k$ for early fusion, or enable mid- or late-fusion. More sophisticated operators (e.g., attention-based fusion) can be integrated as future work. Currently, our framework supports all these variants. Users can also specify the number of principal components for PCA and the number of canonical variables for CCA.
    \item[Model and hyper-parameters:] Select the recommendation backbone (\texttt{cf}, \texttt{vbpr}, \texttt{amr}, \texttt{vmf}, \texttt{vaecf}) and optionally provide model-specific hyper-parameters. The modular design allows researchers to introduce graph-based or transformer-based recommenders with minimal changes.
    \item[Runtime options:] Toggle fast prototypes (a single training epoch) for quick testing, specify a GPU for hyper-parameter search and set random seeds to ensure reproducibility. Logging options can be extended to record additional metadata or integrate with experiment tracking tools.
\end{description}
This declarative configuration approach reduces boilerplate code, ensures that experiments are reproducible, and eases the extension of the benchmark to new domains. 

\subsection{Textual Data Enrichment and Embedding.}
Given an item \(i\) described by title, genre list, and user tags, we generate a canonical text view \(\mathcal{T}(i)\) in one of two mutually exclusive modes:

\textbf{(1) No Augmentation (\textsc{NA}):} After lower-casing and removing structural delimiters, we concatenate the following:

\[
  \mathcal{T}_{\mathrm{NA}}(i)=
  \texttt{title}_i\;+\;
  \text{`` ''}\;+\;
  \texttt{genres}_i\bigl[\!\text{`|'}\!\mapsto\!\text{` '}\bigr]\;+\;
  \text{`` ''}\;+\;
  \bigl(\texttt{tags}_i\;\text{space-joined}\bigr).
\]

\textbf{(2) LLM-based Augmentation (\textsc{A}):}  
If synopses are missing, sparse, or inconsistent, a large language model (LLM) is prompted once per item as follows:

\begin{tcolorbox}[colback=gray!5!white, colframe=black!60!black, title=Synopsis Generation Template]
\textbf{Role:} You are a helpful assistant.

\textbf{Task:} Write a vivid, engaging \textbf{100--150-word} synopsis for a movie or artist.

\textbf{Inputs:}
\begin{itemize}
    \item \textbf{Title}: \emph{[Movie/Artist Title]}
    \item \textbf{Genre List}: \emph{[List of genres, e.g., drama, mystery, thriller]}
    \item \textbf{Tags}: \emph{[Comma-separated, free-form tags, e.g., coming-of-age, family, 1980s]}
\end{itemize}

\end{tcolorbox}

Passing \texttt{title}$_i$, \texttt{genres}$_i$, and \texttt{tags}$_i$ as the \texttt{user} message yields the enriched synopsis \(\mathcal{T}_{\mathrm{A}}(i)\), which is stored verbatim.

Regardless of the mode, the resulting text is embedded using a configurable model (\emph{OpenAI-Ada}, \emph{Sentence-T5}, or fine-tuned \emph{LLaMA-2}):

\[
  \mathbf e_i^{(\mathrm{txt})} = \Phi_{\text{mdl}}\bigl(\mathcal{T}_{\mathrm{NA/A}}(i)\bigr) \in \mathbb{R}^{d_\text{txt}}.
\]

The process---augmentation, tokenisation, batching, and embedding---is fully automated, and documented in \texttt{data\_augment\_llm.ipynb}.

\paragraph{Multimodal Alignment and Fusion.}
Audio, visual, and text keys are intersected to ensure every item has complete features. Three deterministic operators are provided:
\[
\small
\begin{array}{ll}
\textsc{concat}: & \mathbf e_i = [\mathbf e_i^{(\mathrm{aud})};\mathbf e_i^{(\mathrm{vis})};\mathbf e_i^{(\mathrm{txt})}] \\
\textsc{pca}_\rho: & \mathbf e_i = \mathbf P_\rho^\top\,\mathrm{zscore}(\mathbf e_i^{\textsc{concat}}) \\
\textsc{cca}_k: & \mathbf e_i = [\mathbf W_1^\top\mathbf e_i^{(1)}]_{1:k}
\end{array}
\]
Fused vectors are transparently wrapped for use with Cornac.

\paragraph{Data Splitting, Training, and Evaluation.}
Splitting strategies (\textsc{random}, \textsc{temporal}, \textsc{per\_user}) preserve chronology in the test fold. Each recommender is trained using fused features and hyperparameters, with performance tracked across a suite of metrics: recall, nDCG, cold-rate, coverage, novelty, diversity, and calibration bias (see below). Ensemble methods (Borda, RRF, etc.) can be enabled post-hoc without retraining.

\begin{table}[h]
\centering
\caption{Final dataset characteristics after merging with \textsc{MovieLens-1M} and \textsc{MMTF-4K}. Although approximately 3,000 movies were initially augmented, the final item count is lower due to overlap with these datasets.}
    \begin{tabular}{l|c}
    \toprule
    \textbf{Metric} & \textbf{Value} \\
    \midrule
    Total Interactions ($|R|$) & 632,397 \\
    Number of Users ($|U|$) & 6,040 \\
    Number of Items ($|I|$) & 992 \\
    Avg.\ Ratings per User ($|R|/|U|$) & 104.70 \\
    Avg.\ Ratings per Item ($|R|/|I|$) & 637.50 \\
    Sparsity ($|R|/(|U| \cdot |I|)$) & 0.1055\% \\
    \bottomrule
    \end{tabular}
\label{tbl:char}
\end{table}

\subsection{Evaluaion in ViLLA-MMBench }
\label{subsec:metrics}

ViLLA-MMBench evaluates recommendation quality along multiple dimensions. Besides Recall@\emph{K} and nDCG@\emph{K}, the framework computes:
\begin{itemize}
    \item \textbf{Cold-start rate}—the proportion of users or items in the top-\emph{K} that were unseen during training.
    \item \textbf{Coverage}—the fraction of the item catalogue that appears in at least one user’s top-\emph{K} list.
    \item \textbf{Novelty}—the mean negative log-popularity of recommended items, thereby encouraging less popular content.
    \item \textbf{Intra-list diversity}—the mean cosine distance between pairs of recommended items for each user.
    \item \textbf{Calibration bias}—the difference between attribute distributions in recommendations and those in the user’s historical interactions.
\end{itemize}
These metrics can be extended to cover fairness or serendipity in subsequent work. Full experiments typically take two to six hours on a free Colab GPU; local execution is supported through the Python package to avoid the version conflicts previously noted by reviewers.
\subsection{Implementation}

ViLLA-MMBench is implemented in \textbf{Python 3.10} and is designed to provide flexibility, modularity, and reproducibility in one framework.
It supports both CPU and GPU execution scenarios, which can be easily toggled via the configuration file, along with options for CPU parallelization to accelerate training or evaluation.
We provide a complete \texttt{setup.py}-based installation, making it straightforward to install all dependencies locally.
For recommendation tasks, the framework integrates with Cornac \cite{cornac}, offering a robust backend for collaborative filtering and multimodal models.
While the codebase is structured to run locally, either directly or through containerized environments, we also offer a dedicated Google Colab implementation that mirrors the full pipeline, allowing users to benefit from Colab’s GPU resources without setup overhead.
Additionally, we provide a secondary Colab file that demonstrates how to load and call the local Python modules and GitHub repository of the framework directly from within a Google Colab environment, further simplifying reproducibility and experimentation.

\section{Evaluation and Benchmarking}
\label{sec_eval}
In this section, we present the experimental results obtained using the proposed framework, benchmarking movie recommendation performance. Our results provide a comprehensive view of the impact of incorporating both visual and audio features from movie trailers, together with LLM-augmented textual data, on downstream movie recommendation tasks. The experiments are designed to address the following research questions.


\begin{enumerate}[label=\textbf{RQ\arabic*.}, wide=0pt, leftmargin=*]

\item \textbf{Impact of text augmentation with LLMs.}\\[2pt]
What is the impact of \textbf{text augmentation} with Large Language Models (LLMs) on video recommendation performance, particularly measured by Recall@10 and NDCG@10, across selected recommendation models (AMR and VBPR)?

\item \textbf{Modality impact.}
\begin{enumerate}[label=\textbf{2.\alph*.}, wide=0pt, leftmargin=*]

\item Which individual modalities (text, vision, audio) or multimodal combinations are consistently beneficial—or detrimental—for overall performance, as measured by the AUC metrics corresponding to NDCG@10 versus ColdRate@10, and Coverage@10?

\item \textbf{Universality of features.}\\[2pt]
Do any features behave in a ``universal'' manner consistently benefiting or harming both AMR and VBPR backbones, irrespective of the evaluation metric used?

\item \textbf{Projection schemes comparison (CCA vs PCA).}\\[2pt]
Between the two experimented projection (dimensionality-reduction) schemes—95\% PCA and 40-dimensional CCA-- does one clearly dominate the other across evaluation metrics and recommendation backbones, or should the projection choice remain a tunable hyperparameter?

\end{enumerate}

\setcounter{enumi}{2}
\item \textbf{Model-based fusion impact.}\\[2pt]
What is the impact of \textbf{model-based fusion approaches} when combining collaborative filtering with multimodal models?

\end{enumerate}

\begin{figure}[t]
    \centering
    \begin{subfigure}[t]{.49\columnwidth}
        \includegraphics[width=\linewidth]{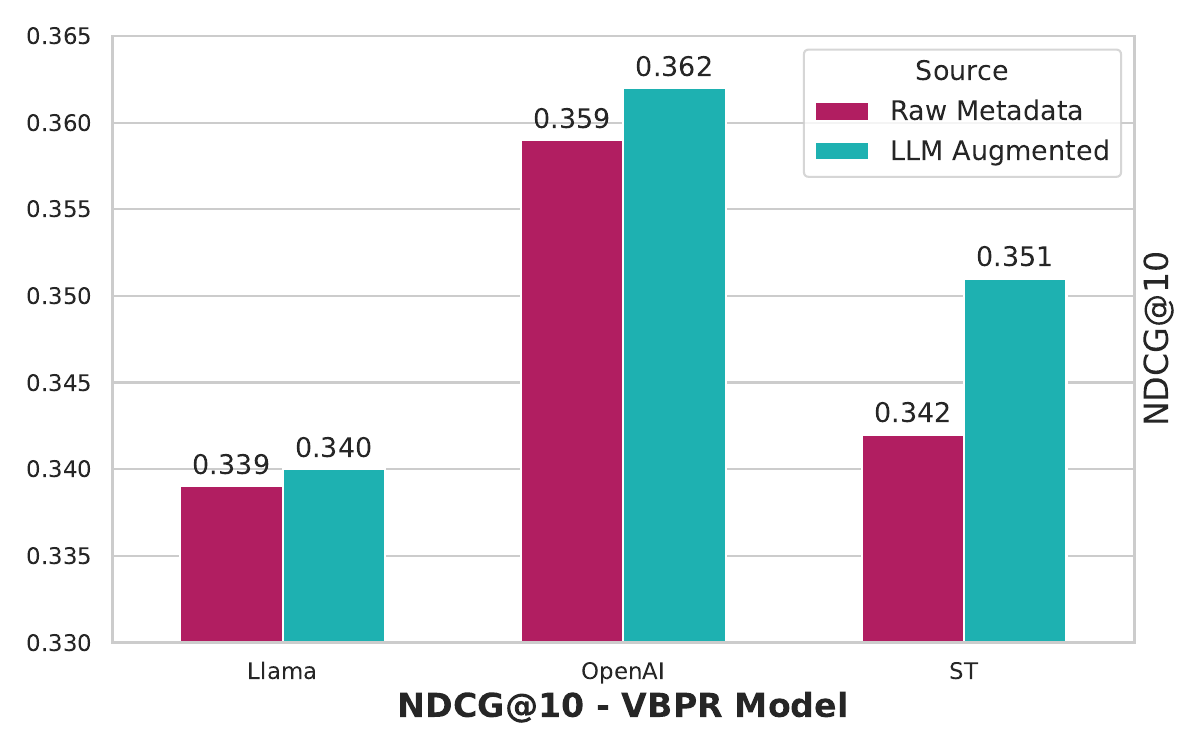}
    \end{subfigure}
    \hspace{0.2mm}
    \begin{subfigure}[t]{.49\columnwidth}
        \includegraphics[width=\linewidth]{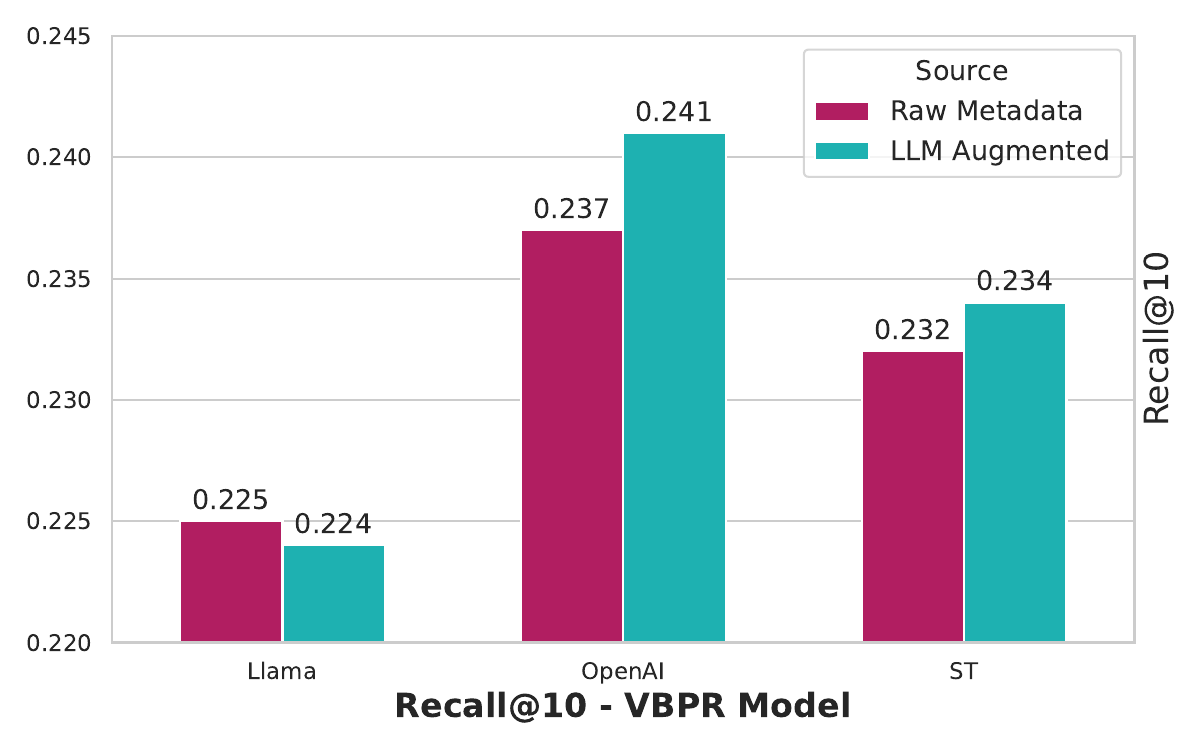}
    \end{subfigure}
    \hspace{0.25mm}
    \begin{subfigure}[t]{.49\columnwidth}
        \includegraphics[width=\linewidth]{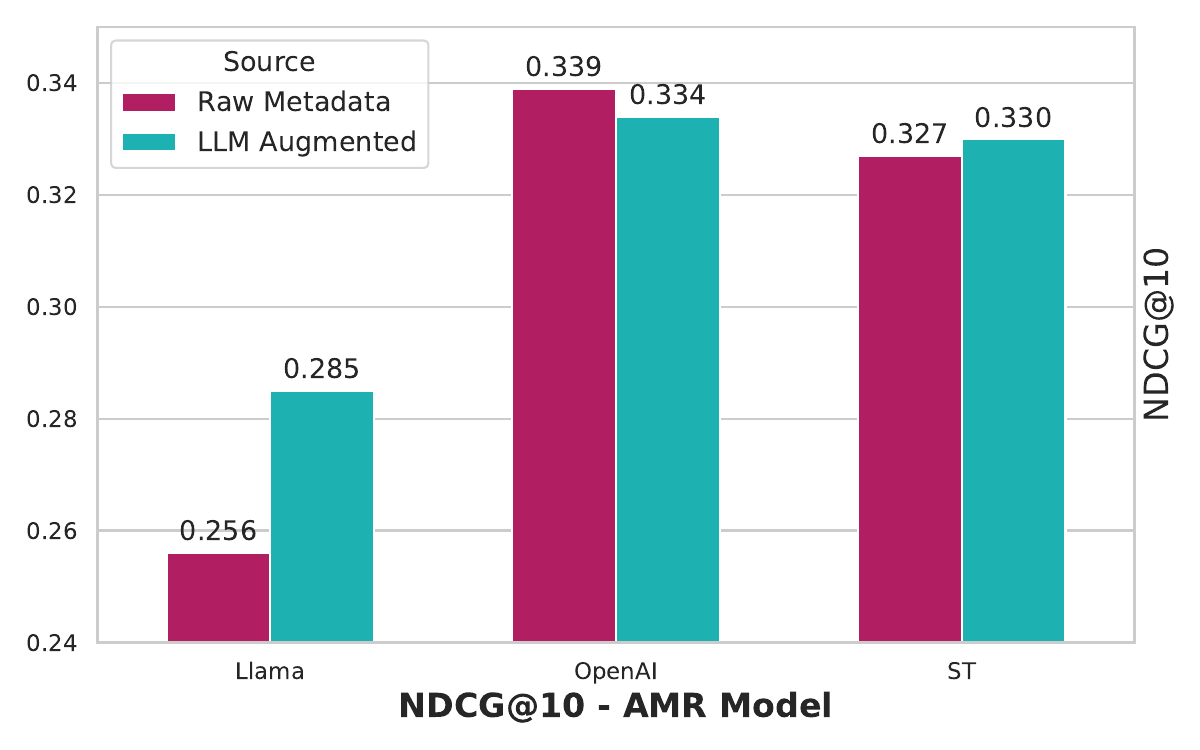}
    \end{subfigure}
    \hspace{0.25mm}
    \begin{subfigure}[t]{.49\columnwidth}
        \includegraphics[width=\linewidth]{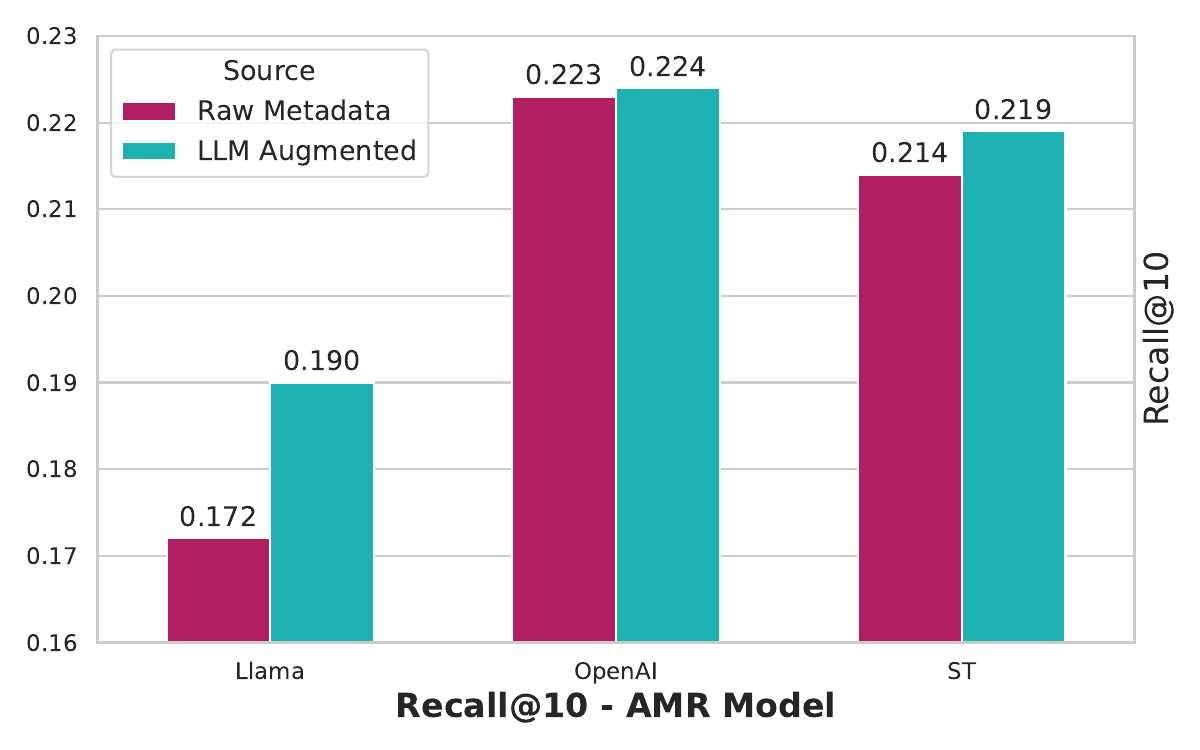}
    \end{subfigure}
    \caption{Effect of LLM-based text augmentation on NDCG@10 and Recall@10 for unimodal textual sources.}
    \label{fig:eval_text_aug}
\end{figure}

\subsection{RQ1. (LLM-based text augmentation Impact)}

Textual metadata for video items is often sparse or noisy, limiting the effectiveness of text-aware recommenders. Augmenting this metadata with Large Language Models (LLMs) can enrich semantic representations, potentially improving recommendation quality despite risks of added noise or fairness issues. We present an initial analysis within our framework, encoding augmented text using dense embeddings (LLaMA, OpenAI, \& SentenceTransformer).

A key question we address is \textit{how sentence-level embeddings interact differently with the two recommender backbones?} Results summarizing our observations are shown in Fig.~\ref{fig:eval_text_aug} and detailed below:

\begin{itemize}[leftmargin=*]

    \item \textbf{SentenceTransformer (ST)} embeddings consistently yield modest but reliable improvements for both backbones and metrics:
    Recall increases by \textbf{+0.9\%} on VBPR (0.232~$\rightarrow$~0.234) and by \textbf{+2.3\%} on AMR (0.214~$\rightarrow$~0.219); NDCG rises by \textbf{+2.6\%} on VBPR (0.342~$\rightarrow$~0.351) and by \textbf{+0.9\%} on AMR (0.327~$\rightarrow$~0.330).
    This indicates that ST effectively captures general semantic cues beneficial across models without specific tuning.

    \item \textbf{OpenAI} embeddings exhibit conservative improvements for VBPR—Recall \textbf{+1.7\%}, NDCG \textbf{+0.8\%}—but produce mixed results with AMR: minimal Recall improvement (\textbf{+0.4\%}) accompanied by a slight NDCG decrease (\textbf{--1.5\%}, 0.339~$\rightarrow$~0.334). 
    Hence, OpenAI embeddings remain a reliable choice for VBPR but might slightly degrade re-ranking effectiveness by AMR.

    \item \textbf{LLaMA} embeddings show strong model-specific effects: neutral on VBPR (Recall \textbf{--0.4\%}, NDCG \textbf{+0.3\%}), yet highly beneficial for AMR with pronounced gains in Recall (\textbf{+10.5\%}, 0.172~$\rightarrow$~0.190) and NDCG (\textbf{+11.3\%}, 0.256~$\rightarrow$~0.285).
    This suggests that embedding by LLamA structure aligns particularly well with AMR's item-aware layers, providing limited added value to VBPR.

\end{itemize}

\noindent In summary, ST embeddings serve as a robust, universal baseline; OpenAI embeddings provide consistent performance for VBPR but might require further tuning for AMR; LLaMA embeddings offer significant benefits for AMR, yet minimal advantage or slight detriment for VBPR. Summaries of these insights can be found in top part of Table 
\ref{tab:compact_combined_color}.

\begin{table}[h!]
\centering
\renewcommand{\arraystretch}{1.08}
\caption{Top: \% change for each embedding baseline. Bottom: Recommended configs by KPI/backbone.}
\label{tab:compact_combined_color}
\begin{tabular}{@{}lcccc|c@{}}
\toprule
\textbf{Emb.} & \textbf{VB R} & \textbf{VB N} & \textbf{AMR R} & \textbf{AMR N} & \textbf{Avg} \\
\midrule
ST      & +0.9  & +2.6  & +2.3  & +0.9   & \textbf{1.7}  \\
OpenAI  & +1.7  & +0.8  & +0.4  & --1.5  & \textbf{0.4}  \\
Llama   & --0.4 & +0.3  & +10.5 & +11.3  & \textbf{5.4}  \\
\arrayrulecolor{gray!55}\midrule
\rowcolor{gray!18}
\emph{Avg} & \emph{0.7} & \emph{1.2} & \emph{4.4} & \emph{3.6} & \emph{3.0} \\
\arrayrulecolor{black}\bottomrule
\end{tabular}

\vspace{0.45em}

\renewcommand{\arraystretch}{1.04}
\begin{tabular}{@{}ll@{}}
\toprule
\textbf{KPI} & \textbf{Recommended Configs (AMR / VBPR)} \\
\midrule
Cold-start & \small{\textcolor{magenta}{Raw OpenAI} + \textcolor{teal}{CNN} + \textcolor{blue}{BLF} \quad / \quad \textcolor{magenta}{Aug ST (text-only)}} \\
Coverage   & \small{\textcolor{magenta}{Raw OpenAI} + \textcolor{teal}{AVF} + \textcolor{blue}{i-vec} \quad / \quad \textcolor{magenta}{Aug OpenAI} + \textcolor{teal}{AVF} + \textcolor{blue}{i-vec}} \\
\midrule
\multicolumn{2}{l}{\footnotesize \textbf{Tips:} Strong text emb. matter most; \textcolor{teal}{AVF}+\textcolor{blue}{i-vec} boost Coverage.} \\
\bottomrule
\end{tabular}

\vspace{0.1em}
{\footnotesize \textbf{R}: Recall, \textbf{N}: NDCG, VB: VBPR. ST is consistently positive; OpenAI excels on VBPR; Llama excels for AMR.}
\end{table}

\subsection{RQ2-a. (Modality Impact)}
\begin{figure*}[ht]
  \centering
  \includegraphics[width=0.9\linewidth]{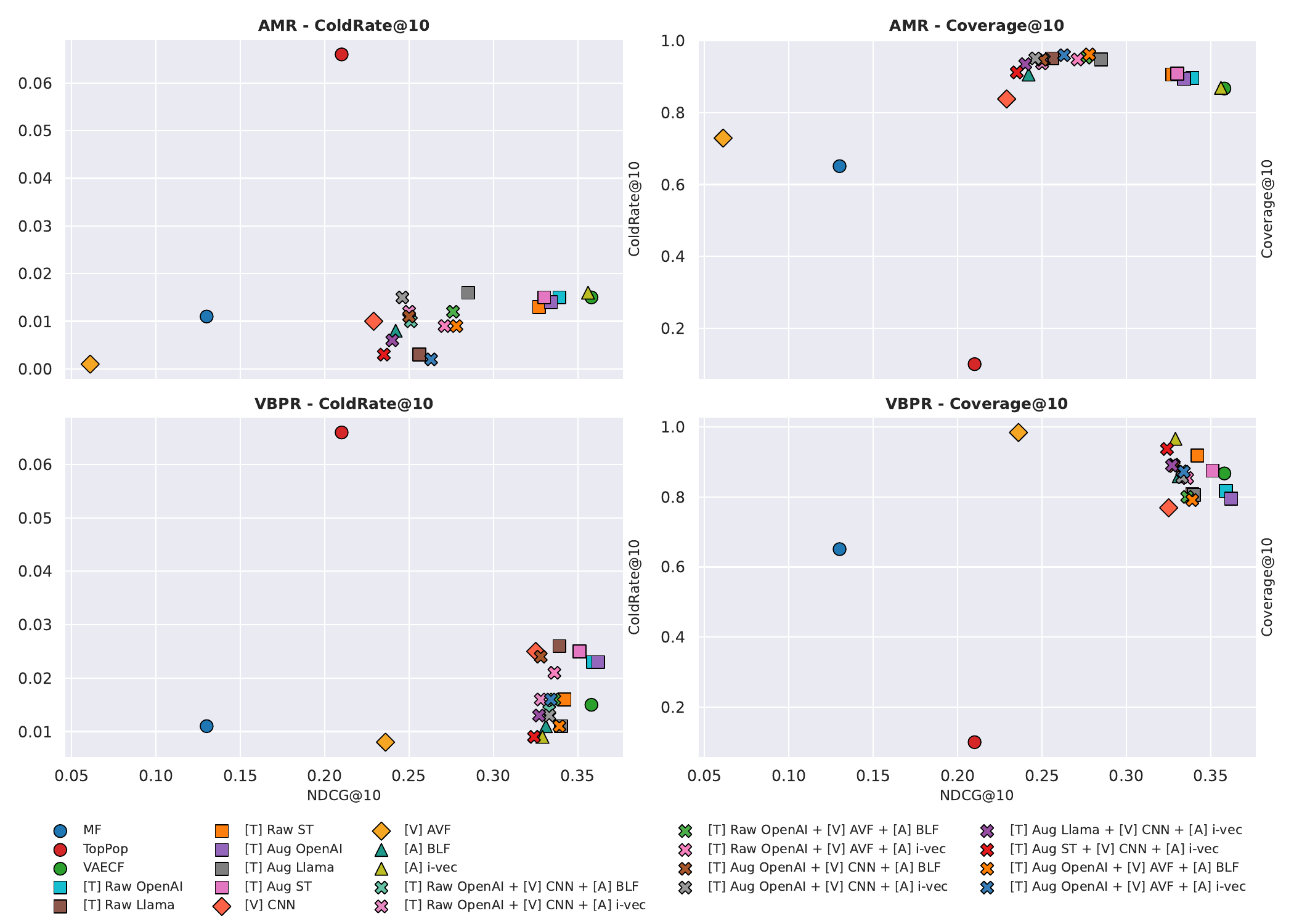}
\caption{Evaluation results across various model and modality combinations. Multimodal data are fused using Principal Component Analysis (PCA) to combine modalities. We provide this visualization for clarity. However, the results using Canonical Correlation Analysis (CCA) are also available at the provided link.}
  \label{fig_eval_combined}
\end{figure*}

Here we aim to analyze the impact of multi-modal features across two recommender backbones, \textbf{AMR} and \textbf{VBPR}, focusing on cold start performance (NDCG-ColdRate@10 AUC) and catalog coverage (NDCG-Coverage@10 AUC). As these metrics are intended to reflect aspects beyond pure accuracy, we ensure our analysis is grounded in systems that already perform well in terms of ranking quality—measured by NDCG. Therefore, we base the following discussion on AUC values derived from the NDCG-ColdRate and NDCG-Coverage curves (after min-max normalizing the values per model VBPR and AMR). The raw values are reported in Figure \ref{fig_eval_combined}. For brevity, we refer to the AUC of NDCG-ColdRate@10 as \textit{ColdRate@10 AUC}, and the AUC of NDCG-Coverage@10 simply as \textit{Coverage@10 AUC} in the following discussion. \\

\vspace{-1mm}

\noindent \textbf{Note.} We use \textit{Raw} for original text and \textit{Aug} for LLM-generated text. Since RQ1 shows the advantage of augmentation, most multimodal combinations use the augmented version for the effort consideration.


\subsubsection{AMR Backbone}
For \textbf{AMR}, distinct modality combinations optimize each KPI differently. For cold-start, the combination of textual (\textcolor{magenta}{Raw OpenAI}), visual (\textcolor{teal}{CNN}), and audio (\textcolor{blue}{BLF}) modalities clearly outperforms other configurations, achieving the highest observed ColdRate@10 AUC of \textbf{0.8576} (NDCG@10 = 0.314, ColdRate@10 = 0.019). Interestingly, using only audio (\textcolor{blue}{i-vec}) also offers good cold-start performance (AUC \textbf{0.8333}, NDCG@10 = 0.356, ColdRate@10 = 0.016), but trails the multimodal \textcolor{teal}{CNN}+\textcolor{blue}{BLF} combo by approximately 2.4 percentage points. Pure-text configurations, while foundational, remain significantly below multimodal runs in cold-start effectiveness—for example, \textcolor{magenta}{Raw OpenAI} alone achieves an AUC of only \textbf{0.7330}, demonstrating a substantial multimodal advantage (+0.12).

In terms of catalog coverage, the optimal AMR configuration shifts distinctly. Here, the combination of textual (\textcolor{magenta}{Raw OpenAI}), aesthetic visual features (\textcolor{teal}{AVF}), and audio (\textcolor{blue}{i-vec}) using a CCA projection yields the highest AUC (\textbf{0.7452}, with NDCG@10 = 0.321 and Coverage@10 = 0.926). Notably, this clearly surpasses pure-text solutions like \textcolor{magenta}{Aug ST} (AUC \textbf{0.7044}, NDCG@10 = 0.33, Coverage@10 = 0.909), highlighting the importance of multimodality for effectively exploring the long tail issue of the catalog.

\subsubsection{VBPR Backbone}
For \textbf{VBPR}, however, the modality effects differ substantially. Cold-start performance strongly favors textual modalities alone. Specifically, \textcolor{magenta}{Augmented SentenceTransformer (Aug ST)} achieves the highest ColdRate@10 AUC (\textbf{0.8620}), closely followed by other text-only methods (\textcolor{magenta}{Aug OpenAI}, \textbf{0.8333}, and \textcolor{magenta}{Raw Llama}, \textbf{0.8175}). Incorporating visual or audio modalities significantly deteriorates cold-start effectiveness, exemplified by the visual modality alone (\textcolor{teal}{CNN}), which yields an AUC of only \textbf{0.6671} (NDCG@10 = 0.325, ColdRate@10 = 0.025). Thus, cold-start effectiveness by VBPR hinges exclusively on textual embeddings.

For catalog coverage, however, VBPR mirrors AMR in preferring multimodal approaches. The best-performing VBPR coverage configuration again includes textual (\textcolor{magenta}{Aug OpenAI}), visual (\textcolor{teal}{AVF}), and audio (\textcolor{blue}{i-vec}) modalities, achieving the highest Coverage@10 AUC of \textbf{0.7051}. The NDCG@10 and Coverage@10 values by this configuration are 0.336 and 0.96, respectively. Pure text modalities remain substantially behind (AUC $\leq$ \textbf{0.58}), underscoring the critical role of multimodal fusion to maximize catalog exploration.

\vspace{0.5em}
\subsection{RQ2-b. (Universality)}

We examined whether specific features consistently enhance or impair performance across both backbones and metrics as shown in Table ~\ref{tab:compact_combined_color}:\\

\textbf{Universally beneficial:} 
Strong textual embeddings, particularly from large language models like \textcolor{magenta}{OpenAI} or \textcolor{magenta}{Aug ST}, consistently form the foundation of high-performing configurations across both backbones and metrics. \\

\textbf{Nearly universally beneficial (for Coverage):}
The fusion of visual aesthetics (\textcolor{teal}{AVF}) with audio i-vectors (\textcolor{blue}{i-vec}) consistently enhances catalog coverage performance, as demonstrated by their presence in top-ranking configurations for both AMR (AUC = \textbf{0.7452}) and VBPR (AUC = \textbf{0.7051}). \\

\textbf{Model-specific modality effects:}
The combination of visual CNN features (\textcolor{teal}{CNN}) with audio block-level features (\textcolor{blue}{BLF}) significantly benefits AMR cold-start performance but consistently harms VBPR cold-start performance. This highlights that certain modality combinations should be specifically tailored to the recommender backbone used. 

\subsection{RQ2-c. (Projection Schemes)}

We analyzed two projection schemes—Principal Component Analysis (\textbf{PCA-95}) and Canonical Correlation Analysis (\textbf{CCA-40})—to understand their relative benefits for multimodal recommender performance across AMR and VBPR backbones, considering AUC of both ColdRate@10 and Coverage@10 metrics. 

\subsubsection{AMR Backbone}

For the AMR backbone, CCA substantially outperformed PCA in both evaluation metrics, clearly dominating the performance landscape. In the cold-start scenario, the top-performing CCA-based configuration (\textcolor{magenta}{Raw OpenAI} + \textcolor{teal}{CNN} + \textcolor{blue}{BLF}) achieved an exceptionally high AUC of \textbf{0.8576}, vastly exceeding the best PCA-based configuration (\textcolor{magenta}{Aug OpenAI} + \textcolor{teal}{CNN} + \textcolor{blue}{i-vec}) at only \textbf{0.4878}—a remarkable absolute improvement. Such a substantial gap indicates that the cross-modal alignment captured by CCA is essential to effectively handle cold-start challenges posed by AMR.

Regarding Coverage@10, CCA again provided the superior choice, albeit with a smaller but still meaningful margin. The top CCA-based model (\textcolor{magenta}{Raw OpenAI} + \textcolor{teal}{AVF} + \textcolor{blue}{i-vec}) reached an AUC of \textbf{0.7452}, surpassing PCA’s best configuration (\textcolor{magenta}{Aug OpenAI} + \textcolor{teal}{AVF} + \textcolor{blue}{BLF}) with an AUC of \textbf{0.7356}. Although the margin is smaller (+0.0096), the consistent advantage reinforces CCA’s suitability as the default projection method for AMR.

\subsubsection{VBPR Backbone}

The VBPR backbone, however, exhibited a mixed picture. For the ColdRate@10 metric, PCA showed a clear advantage. Specifically, leading configuration in PCA (\textcolor{magenta}{Aug OpenAI} + \textcolor{teal}{CNN} + \textcolor{blue}{BLF}) attained an AUC of \textbf{0.6490}, significantly above the best-performing CCA configuration (\textcolor{magenta}{Aug OpenAI} + \textcolor{teal}{CNN} + \textcolor{blue}{i-vec}) which only reached \textbf{0.3968}. This indicates the ability of PCA to preserve the distinctive modality-specific variance essential for VBPR’s cold-start performance.

In contrast, when assessing Coverage@10, CCA regained dominance. The highest performing configuration under CCA (\textcolor{magenta}{Aug OpenAI} + \textcolor{teal}{AVF} + \textcolor{blue}{i-vec}) scored \textbf{0.7051}, substantially outperforming leading configuration offered by PCA, (\textcolor{magenta}{Aug ST} + \textcolor{teal}{CNN} + \textcolor{blue}{i-vec}) at \textbf{0.5457}. This difference of approximately 16 percentage points emphasizes the strength of CCA in distributing recommendations more broadly across the catalog.

These results provide insightful guidance for future modeling. For AMR-based recommenders, CCA is the preferred projection method, consistently improving cold-start performance and coverage. For VBPR, PCA is optimal when prioritizing cold-start novelty, while CCA is better for catalog coverage or fairness.

\begin{table}[htbp]
    \centering
    \caption{Comparison of model-based fusion (VAECF + AMR variants). Best per block in \textbf{bold}.}
    \label{tab:fusion_methods_compact}
    \begin{tabular}{l|l|ccc}
    \toprule
        \textbf{Fusion} & \textbf{Model} & \textbf{Rec@10} & \textbf{N@10} & \textbf{HR@10} \\
    \midrule
        \multicolumn{5}{l}{\emph{Audio (AMR Audio)}} \\
        RRF        & VAECF + AMR Audio  & \textbf{.2521} & \textbf{.3739} & .8610 \\
        Borda      & VAECF + AMR Audio  & .2517 & .3732 & \textbf{.8615} \\
        Avg Rank   & VAECF + AMR Audio  & .2520 & .3738 & .8606 \\
    \midrule
        \multicolumn{5}{l}{\emph{Text (AMR Text, OpenAI)}} \\
        RRF        & VAECF + AMR Text \scriptsize{(OpenAI)} & \textbf{.2496} & \textbf{.3666} & \textbf{.8605} \\
        Borda      & VAECF + AMR Text \scriptsize{(OpenAI)} & .2496 & .3647 & \textbf{.8605} \\
        Avg Rank   & VAECF + AMR Text \scriptsize{(OpenAI)} & .2486 & .3661 & .8593 \\
    \midrule
        \multicolumn{5}{l}{\emph{Text (AMR Text, ST)}} \\
        RRF        & VAECF + AMR Text \scriptsize{(ST)} & \textbf{.2490} & \textbf{.3698} & .8564 \\
        Borda      & VAECF + AMR Text \scriptsize{(ST)} & .2490 & .3682 & .8564 \\
        Avg Rank   & VAECF + AMR Text \scriptsize{(ST)} & .2486 & .3690 & \textbf{.8568} \\
    \midrule
        \multicolumn{5}{l}{\emph{No Fusion}} \\
        ---        & VAECF Only & .2492 & .3584 & .8581 \\
    \bottomrule
    \end{tabular}
    \vspace{0.4em}
    
    {\footnotesize
    Rec: Recall, N: NDCG, HR: HitRate, ST: SentenceTransformer.
    }
\end{table}

\vspace{0.5em}

\subsection{RQ3. (Model-based Fusion Impact)}
\label{eval_rq3}

Table~\ref{tab:fusion_methods_compact} summarizes the comparative results. Note that our primary goal here is to combine the best-performing CF model (VACEF) with the best multimodal model to further enhance performance. Rank-based fusion (esp. RRF and Borda) consistently improves performance over single-model VAECF, across both audio and text modalities. Gains are most visible in NDCG and Recall, and Hit Rate. This confirms the value of model-based fusion and rank aggregation methods for leveraging diverse modality-specific signals in video recommendation.

\section{Conclusion}
\label{sec_conclusion}

The proposed toolkit, \textbf{ViLLA-MMBench}, offers a lightweight, reproducible framework that elevates the classic MovieLens benchmark into a comprehensive multi-modal testbed for recommendation research. By combining visual features from trailers, audio embeddings, and LLM-generated text, the toolkit systematically evaluates both individual and fused modalities across a broad spectrum of metrics, including not only accuracy but also beyond-accuracy criteria such as cold-start handling, fairness, novelty, diversity, and catalog coverage.

A distinguishing contribution of \textbf{ViLLA-MMBench} is the automated augmentation of sparse or missing item metadata using state-of-the-art Large Language Models (LLMs), specifically OpenAI’s GPT, which enables the generation of high-quality synopses and consistent textual signals for every movie. Multiple dense embedding types—including OpenAI Ada, LLaMA-2, Sentence-T5, CNN, AVF, BLF, and i-vector—are aligned and made available, supporting interchangeable early-, mid-, and late-fusion strategies and facilitating principled ablation studies.

The fully scripted and logged pipeline, driven by declarative YAML configuration, ensures transparency, repeatability, and ease of extension to new modalities, recommendation backbones, or evaluation protocols. With robust support for MovieLens (100K/1M), MMTF-14K, and a custom LLM-augmented review dataset, as well as modular interfaces for integrating additional data sources, \textbf{ViLLA-MMBench} provides a solid foundation for rigorous benchmarking and fair comparison in multimodal recommendation.

Empirical results demonstrate clear improvements in cold-start and catalog coverage, particularly in scenarios where LLM-augmented text is fused with audio-visual descriptors. In summary, via making all code, embeddings, and configuration templates openly available, this toolkit aims to foster reproducible, extensible, and responsible research, paving the way for principled integration of generative AI in recommender systems. Future work will explore further modalities, additional domains, and more advanced evaluation criteria, continuing to advance the state of the art in trustworthy, multi-modal recommendations.



\bibliographystyle{IEEEtran}
\bibliography{refs}

\end{document}